\documentclass[11pt,a4paper]{article}

\usepackage[]{hyperref}
\usepackage{amsmath, amsthm, amssymb}
\usepackage{graphicx}

\def\ket#1{\left|#1\right>}

\title{The measurement problem revisited}

\author{Shan Gao
\\Research Center for Philosophy of Science and Technology, 
\\ Shanxi University, Taiyuan 030006, P. R. China
\\ E-mail:  \href{mailto:gaoshan2017@sxu.edu.cn}{gaoshan2017@sxu.edu.cn}.}

\begin{document}
\maketitle

\begin{abstract}\noindent 

It has been realized that the measurement problem of quantum mechanics is essentially the determinate-experience problem, and in order to solve the problem, the physical state representing the measurement result is required to be also the physical state on which the mental state of an observer supervenes. This necessitates a systematic analysis of the forms of psychophysical connection in the solutions to the measurement problem. 
In this paper, I propose a new, mentalistic formulation of the measurement problem which lays more stress on psychophysical connection. 
By this new formulation, it can be seen more clearly that the three main solutions to the measurement problem, namely Everett's theory, Bohm's theory and collapse theories,  correspond to three different forms of psychophysical connection. I then analyze these forms of psychophysical connection. 
It is argued that the forms of psychophysical connection required by Everett's and Bohm's theories have potential problems, while an analysis of how the mental state of an observer supervenes on her wave function may help solve the structured tails problem of collapse theories. 
\end{abstract}

\vspace{6mm}

\section{Introduction}

The measurement problem is a long-standing problem of quantum mechanics.
The theory assigns a wave function to an isolated physical system and specifies that the evolution of the wave function is governed by the Schr\"{o}dinger equation. However, when assuming the wave function is a complete description of the system, the linear dynamics is apparently incompatible with the appearance of definite results of measurements on the system. This leads to the measurement problem. 
Maudlin (1995a) gave a precise formulation of the problem in terms of the incompatibility.
Correspondingly, the three approaches to avoiding the incompatibility lead to the three main solutions to the measurement problem: Everett's theory, Bohm's theory and collapse theories. 
It is widely thought that these theories can indeed solve the measurement problem, although each of them still has some other problems.

On the other hand, it has been realized that the measurement problem of quantum mechanics is essentially the determinate-experience problem  (Barrett, 1999).
In the final analysis, the problem is to explain how the linear dynamics can be compatible with the existence of definite experiences of conscious observers. 
This requires that in the above solutions to the measurement problem the physical state representing the measurement result should be also the physical state on which the mental state of an observer supervenes.\footnote{In this paper, supervenience will be always used in its standard definition. A set of properties $A$ supervenes on another set $B$ in case no two things can differ with respect to $A$-properties without also differing with respect to their $B$-properties (see McLaughlin and Bennett, 2014). By this definition, the principle of psychophysical supervenience requires that the mental properties of a system cannot change without the change of its physical properties. 
In this paper I will not consider the possibility that there is no physical state representing the measurement result on which the mental state of an observer may supervene.} 
As a result, different solutions to the measurement problem may correspond to different forms of psychophysical connection. 
However, this aspect of the measurement problem is ignored in Maudlin's (1995a) formulation. Moreover, 
although there have been some interesting analyses of psychophysical connections in the three main solutions to the measurement problem (Albert, 1992; Brown, 1996; Butterfield, 1998; Barrett, 1999; Brown and Wallace, 2005; Lewis, 2007a), these analyses seems still not complete. 
In this paper, I will propose a new, mentalistic formulation of the measurement problem which gives prominence to the psychophysical connection, and present a new analysis of the forms of psychophysical connection required by the three main solutions to the measurement problem. In particular, I will analyze  whether each form of psychophysical connection satisfies the principle of psychophysical supervenience and how the mental state of an observer supervenes on her wave function.  

%

This paper is organized as follows. 
In Section 2, I first introduce Maudlin's conventional formulation of the measurement problem, and then suggest a new,  mentalistic formulation of  the problem which lays more stress on the aspect of psychophysical connection. 
It is pointed out that the three main solutions to the measurement problem, namely Everett's theory, Bohm's theory and collapse theories, correspond to three different forms of psychophysical connection. 
In Section 3, Everett's theory is analyzed.
The theory requires that the mental state of an observer is not always determined by her whole wave function, and especially, for a post-measurement wave function there are many mental states, each of which supervenes on a certain branch of the wave function. 
It is argued that this form of psychophysical connection seems to violate psychophysical supervenience.  
In Section 4, Bohm's theory is analyzed. 
It is argued that the two suggested forms of psychophysical connection of the theory both have potential problems. In particular, the well-accepted form of psychophysical connection (i.e. the form that the mental state of an observer supervenes on the configuration of her Bohmian particles) may lead to the problem of allowing superluminal signaling. 
In Section 5, I analyze collapse theories, in particular, how the mental state of an observer supervenes on her wave function in these theories. It is argued that the analysis may help solve the structured tails problem of  collapse theories. 
Conclusions are given in the last section.

\section{A mentalistic formulation of the measurement problem}

According to Maudlin's (1995a) formulation, the measurement problem originates from the incompatibility of the following three claims:

(C1). the wave function of a physical system is a complete description of the system; \\
\indent (C2). the wave function always evolves in accord with a linear dynamical equation, e.g. the Schr\"{o}dinger equation; \\
\indent  (C3). each measurement has a definite result (which is one of the possible measurement results whose probability distribution satisfies the Born rule). 

The proof of the inconsistency of these three claims is familiar. 
Suppose a measuring device $M$ measures the $x$-spin of a spin one-half system $S$ that is in a superposition of two different $x$-spins $1/\sqrt{2}(\ket{up}_S+\ket{down}_S)$. If (C2) is correct, then the state of the composite system after the measurement must evolve into the superposition of $M$ recording $x$-spin up and $S$ being $x$-spin up and $M$ recording $x$-spin down and $S$ being $x$-spin down:

\begin{equation}
1/\sqrt{2}(\ket{up}_S \ket{up}_M+\ket{down}_S \ket{down}_M).
\label{ds}
\end{equation}

\noindent The question is what kind of state of the measuring device this represents.
If (C1) is also correct, then this superposition must specify every physical fact about the measuring device. But by symmetry of the two terms in the superposition, this superposed state cannot describe a measuring device recording either $x$-spin up or $x$-spin down. Thus if (C1) and (C2) are correct, (C3) must be wrong.

It can be seen that there are in general three approaches to solving the measurement problem thus formulated.
The first approach is to deny the claim (C1), and add some additional variables and corresponding dynamics to explain the appearance of definite measurement results. A well-known example is Bohm's theory (Bohm, 1952).
The second approach is to deny the claim (C2), and revise the Schr\"{o}dinger equation by adding some nonlinear and stochastic evolution terms to explain the appearance of definite measurement results. Such theories are called collapse theories (Ghirardi, 2011).
The third approach is to deny the claim (C3), and assume the existence of many equally real worlds to accommodate all possible results of measurements (Everett, 1957; DeWitt and Graham, 1973). In this way, it may also explain the appearance of definite measurement results in each world including our own world. This approach is called Everett's interpretation of quantum mechanics or Everett's theory.

It has been realized that the measurement problem in fact has two levels: the physical level and the mental level, and it is essentially the determinate-experience problem (Barrett, 1999). 
The problem is not only to explain how the linear dynamics can be compatible with the appearance of definite measurement results obtained by physical devices, but also, and more importantly, to explain how the linear dynamics can be compatible with the existence of definite experiences of conscious observers. 
However, the mental aspect of the measurement problem is ignored in Maudlin's (1995a) formulation, which defines the problem at the physical level.\footnote{It seems that Maudlin's ignorance is deliberate and he also has a defense for it (Maudlin, 2007). According to Maudlin (2007), we had better avoid explaining how determinate conscious experiences supervene on the wave function, since this brings in the mind-body problem, the problem of explaining how consciousness could supervene on anything physical in the first place, a problem which many take to be unsolvable. My response is that if there is no any psychophysical connection, the measurement problem, which is essentially the determinate-experience problem, cannot even be  formulated at the physical level as Maudlin (1995a) did. In other words, Maudlin's (1995a) formulation of the measurement problem implicitly assumes a certain form of psychophysical connection (see later discussion).} 
Here I will suggest a new, mentalistic formulation of the measurement problem which defines the problem at the mental level and lays more stress on the psychophysical connection. 
In the formulation, the measurement problem originates from the incompatibility of the following three assumptions:

(A1). the mental state of an observer supervenes on her wave function; \\ 
\indent (A2). the wave function always evolves in accord with a linear dynamical equation, e.g. the Schr\"{o}dinger equation; \\
\indent (A3). a measurement does not branch an observer into multiple observers.


The proof of the inconsistency of these assumptions is similar to the above proof. 
Suppose an observer $M$ measures the $x$-spin of a spin one-half system $S$ that is in a superposition of two different $x$-spins, $1/\sqrt{2}(\ket{up}_S+\ket{down}_S)$. 
If (A2) is correct, then the physical state of the composite system after the measurement will evolve into the superposition of $M$ recording $x$-spin up and $S$ being $x$-spin up and $M$ recording $x$-spin down and $S$ being $x$-spin down:

\begin{equation}
1/\sqrt{2}(\ket{up}_S \ket{up}_M+\ket{down}_S \ket{down}_M).
\label{os}
\end{equation}

\noindent If (A1) and (A3) are also correct, then there will be only one observer $M$ throughout the measurement, and the mental state of the observer $M$ will supervene on this superposed wave function.  Since the mental states corresponding to the physical states  $\ket{up}_M$ and $\ket{down}_M$ differ in their mental content, the observer M being in the superposition (\ref{os}) will have a conscious experience different from the experience of M being in each branch of the superposition by the symmetry of the two branches. In other words, the record that M is consciously aware of is neither $x$-spin up nor $x$-spin down when she is physically in the superposition (\ref{os}). This is inconsistent with experimental observations. Therefore, (A1), (A2) and (A3) are incompatible. 


Since the measurement problem is essentially the determinate-experience problem, this new mentalistic formulation of the measurement problem is more appropriate than Maudlin's original physicalistic formulation of the problem. 
As noted before, the measurement problem cannot be formulated if not assuming a psychophysical connection. What we are sure of is that we as observers obtain a definite result and have a definite mental state after a measurement. But we are not sure of what physical state this mental state corresponds to. 
For example, if the mental state supervenes randomly on one branch of the post-measurement superposition such as (\ref{ds}), as in the single-mind theory (Albert and Loewer, 1988), then the three claims in Maudlin's formulation are not incompatible. One may further argue that Maudlin's formulation also needs to rely on the assumption (A1) in the new formulation if it is a valid formulation of the measurement problem. 
Moreover, if the assumption (A1) is included, the claim (C1) will be redundant. Then, Maudlin's formulation will reduce to the new formulation. 

By this new formulation of the measurement problem, we can look at the three main solutions of the problem from a new angle. 
The solution to the measurement problem must deny either the assumption (A1) or the assumption (A2) or the assumption (A3). 
Denying the assumption (A1) means that the mental state of an observer supervenes not on her wave function  but on other additional variables. This corresponds to Bohm's theory. 
Denying the assumption (A2) is the same as denying the claim (C2), which means that the Schr\"{o}dinger equation must be revised. This corresponds to collapse theories. In this case, the mental state of an observer supervenes on her wave function. 
Similarly, denying the assumption (A3) is the same as denying the claim (C3), which means that a measurement branches an observer into multiple observers. This corresponds to Everett's theory.  
It is worth noting that since in this case a post-measurement wave function corresponds to many observers, the mental state of each observer is not uniquely determined by her whole wave function, and it supervenes on one branch of the wave function. 


To sum up,  the three main solutions to the measurement problem, namely Everett's theory, Bohm's theory and collapse theories,  correspond to three different forms of psychophysical connection. 
In fact, there are only three types of physical states on which the mental state of an observer may supervene, which are (1) the wave function, (2) certain branches of the wave function, and (3) other additional variables.
The question is: Exactly what physical state does the mental state of an observer supervene on? 
It can be expected that an analysis of this question may help solve the measurement problem. 

\section{Everett's theory}

I will first analyze Everett's theory. 
The theory assumes that the wave function of a physical system is a complete description of the system, and the wave function always evolves in accord with the Schr\"{o}dinger equation. 
In order to solve the measurement problem, the theory claims that for the above post-measurement state (\ref{os}) there are two observers, and each of them is consciously aware of a definite record, either $x$-spin up or $x$-spin down.\footnote{Note that in Wallace\index{Wallace, David}'s (2012) latest formulation of Everett's theory\index{Everett's theory} the number of the emergent observers after the measurement is not definite due to the imperfectness of decoherence. My following analysis also applies to this case. } 

There are (at least) three ways of understanding the notion of multiplicity in Everett's theory: (1) measurements lead to multiple worlds at the fundamental level (DeWitt and Graham, 1973), (2) measurements lead to multiple worlds only at the non-fundamental ``emergent" level (Wallace, 2012), and (3) measurements only lead to multiple minds (Zeh, 1981; Albert and Loewer, 1988). 
In either case, for the above post-measurement state (\ref{os}), the mental state of each observer is not uniquely determined by her whole wave function, and it supervenes on one branch of the wave function. 
In the following, I will argue that this form of psychophysical connection seems to violate the principle of psychophysical supervenience (see Gao, 2017 for a more detailed analysis).  

Consider a unitary time evolution operator, which changes the first branch of the superposition (\ref{os}) to its second branch and the second branch to the first branch. It is similar to the NOT gate for a single q-bit, and is permitted by the Schr\"{o}dinger equation in principle. Then after the evolution the superposition does not change. According to Everett's theory, the wave function of a physical system is a complete description of the system. Therefore, the physical properties or physical state of the composite system does not change after the unitary time evolution. 

On the other hand, after the evolution the mental state of each observer which supervenes on the corresponding branch of the superposition will change; the mental state supervening on the first branch will change from being aware of $x$-spin up to being aware of $x$-spin down, and the mental state supervening on the second branch will change from being aware of $x$-spin down to being aware of $x$-spin up.\footnote{If this is not the case, then for other evolution or other post-measurement states such as those containing only one branch of the superposition, the predictions of the theory may be inconsistent with the predictions of quantum mechanics and experience.} 
Then, the mental state of each observer does not supervene on the whole superposition or the physical state of the composite system. Since the mental states of the system are composed of the mental states of the two observers, they do not supervene on the physical state of the system either. 
Therefore, it seems that the psychophysical supervenience is violated by Everett's theory in this example. Note again that supervenience is used here in its standard definition, and the principle of psychophysical supervenience requires that the mental properties of a system cannot change without the change of its physical properties. 

It is worth noting that the validity of this argument is independent of the ways of understanding multiplicity in Everett's theory. It is well known that the many-minds theory violates psychophysical supervenience (Albert and Loewer, 1988; Barrett, 1999), and thus the above result is not new for the theory. But for the many-worlds theory, no matter the worlds are at the fundamental level or only at the non-fundamental ``emergent" level, the above result is new; it seems that a many-worlds theory also violates psychophysical supervenience. 

There are two possible ways to avoid the violation of psychophysical supervenience in the above example. 
The first way is to deny that after the evolution the physical state of the composite system has not changed. This requires that the wave function of a system is not a complete description of the physical state of the system. Obviously, this requirement is not consistent with Everett's theory.  
The second way is to deny that after the evolution the total mental states or mental properties of the composite system have changed. 
For example, one may argue that after the above evolution there remain a mental state corresponding to seeing a spin up result and a mental state corresponding to seeing a spin down result, and thus the total mental states of the composite system have not changed. However, this seems to require that each observer has no trans-temporal identity, while the absence of identities of observers is inconsistent with the predictions of quantum mechanics and experience. If each observer has a trans-temporal identity and her mental state supervenes on the corresponding branch of the superposition, then her mental state will change after the evolution, and thus the total mental states or mental properties of the composite system, which are composed of the mental states of these observers, also change after the evolution.\footnote{By comparison, if for the post-measurement superposition (\ref{os}) there is only one observer whose mental content is composed of seeing a spin up result and seeing a spin down result, then her mental state will not change after the above evolution, and the principle of psychophysical supervenience can be satisfied (see further discussion about collapse theories in Section 5).} 

It is usually thought that if a mental state supervenes on part of a physical state then the mental state also supervenes on the physical state. This is indeed the case in the classical domain. But it may be not the case in the quantum domain, e.g. when the physical state is completely described by the wave function. The reason is that when one branch of a wave funcion is changed, if only the other branch is also changed in a particular way, the whole wave function may be unchanged. Then when a mental state supervenes on one branch of the wave function, it may not supervene on the whole wave function; when the branch of the wave function and the corresponding mental state change, the whole wave function may not change. 
 
Finally, I note that if the above analysis is valid, then the measurement problem can be formulated as the incompatibility of only two assumptions: (A1). the mental state of an observer supervenes on her wave function; and (A2). the wave function always evolves in accord with a linear dynamical equation, e.g. the Schr\"{o}dinger equation. The reason is that if a measurement branches an observer into multiple observers as in Everett's theory, then the mental state of an observer will not supervene on her wave function, and thus the third assumption in the previous mentalistic formulation of the measurement problem can be dropped.

\section{Bohm's theory}

Let us turn to Bohm's theory. 
In this theory, there are two suggested forms of psychophysical connection. 
The first one is that the mental state supervenes on the branch of the wave function occupied by Bohmian particles, and the second one is that the mental state supervenes on the (relative) configuration of Bohmian particles. 

The first form of psychophysical connection has been the standard view until recently, according to which the mental state of an observer being in a post-measurement superposition like (\ref{os}) supervenes on the branch of the superposition occupied by her Bohmian particle\index{Bohm's theory!Bohmian particles in}s. 
Indeed, Bohm initially assumed this form of psychophysical connection. He said: ``the packet entered by the apparatus [hidden] variable...  determines the actual result of the measurement, which the observer will obtain when she looks at the apparatus." (Bohm,\index{Bohm, David} 1952, p.182). 
In this case, the role of the Bohmian particle\index{Bohm's theory!Bohmian particles in}s is merely to select the branch from amongst the other non-overlapping branches of the superposition.\index{Bohm's theory!psychophysical supervenience in}

The first form of psychophysical connection is also called Bohm's result assumption (Brown\index{Brown, Harvey. R.} and Wallace\index{Wallace, David}, 2005), and it has been widely argued to be problematic (Stone\index{Stone, Abraham D.}, 1994; Brown\index{Brown, Harvey. R.}, 1996; Zeh\index{Zeh, H. Dieter}, 1999; Brown\index{Brown, Harvey. R.} and Wallace\index{Wallace, David}, 2005; Lewis,\index{Lewis, Peter J.} 2007a). 
For example, according to Brown\index{Brown, Harvey. R.} and Wallace\index{Wallace, David} (2005), in the general case each of the non-overlapping branches in the post-measurement superposition has the same credentials for representing a definite measurement result as the single branch does in the predictable case (i.e. the case in which the measured system is in an eigenstate of the measured observable). The fact that only one of them carries the Bohmian particle\index{Bohm's theory!Bohmian particles in}s does nothing to remove these credentials from the others, and adding the particles to the picture does not interfere destructively with the empty branches either.\index{Bohm's theory!psychophysical supervenience in}

In my view, the main problem with the first form of psychophysical connection is that the empty branches and the occupied branch have the same qualification to be supervened by the mental state.
Moreover, although it is imaginable that the Bohmian particle\index{Bohm's theory!Bohmian particles in}s may have influences on the occupied branch, e.g. disabling it from being supervened by the mental state, it is hardly conceivable that the Bohmian particle\index{Bohm's theory!Bohmian particles in}s have influences on all other empty branches, e.g. disabling them from being supervened by the mental state. 

In view of the first form of psychophysical connection being problematic, most Bohmians today seem to support the second form of psychophysical connection (Lewis, 2007a), although they sometimes do not state it explicitly (Maudlin, 1995b). 
If assuming this form of psychophysical connection, namely assuming the mental state supervenes on the (relative) configuration of Bohmian particle\index{Bohm's theory!Bohmian particles in}s, then the above problems can be avoided.\index{Bohm's theory!psychophysical supervenience in}
However, it has been argued that this form of psychophysical connection is inconsistent with the popular functionalist approach to consciousness (Brown\index{Brown, Harvey. R.} and Wallace\index{Wallace, David}, 2005; see also Bedard, 1999). The argument can be summarized as follows. If the functionalist assumption is correct, for consciousness to supervene on the Bohmian particle\index{Bohm's theory!Bohmian particles in}s but not the wave function, the Bohmian particle\index{Bohm's theory!Bohmian particles in}s must have some functional property that the wave function do not share. But the functional behaviour of the Bohmian particle\index{Bohm's theory!Bohmian particles in}s is arguably identical to that of the branch of the wave function in which they reside. 

Here one may respond, as Lewis (2007b) did, that all theories must give up some intuitive familiar theses and functionalism is the one that Bohm's theory must give up. However, it has been argued that the second form of psychophysical connection also leads to another serious problem of allowing superluminal signaling\index{superluminal signaling}\index{Bohm's theory!superluminal signaling in} (Brown\index{Brown, Harvey. R.} and Wallace\index{Wallace, David}, 2005; Lewis,\index{Lewis, Peter J.} 2007a). If the mental state supervenes on the positions of Bohmian particle\index{Bohm's theory!Bohmian particles in}s, then an observer can in principle know the configuration of the Bohmian particle\index{Bohm's theory!Bohmian particles in}s in her brain with a greater level of accuracy than that defined by the wave function. This will allow superluminal signaling\index{superluminal signaling} and lead to a violation of the no-signaling theorem (Valentini\index{Valentini, Antony}, 1992).

The above analysis of psychophysical supervenience also raises a general doubt about the whole strategy of Bohm's theory\index{Bohm's theory} to solve the measurement problem\index{measurement problem}. Why add hidden variables such as positions of Bohmian particle\index{Bohm's theory!Bohmian particles in}s to quantum mechanics? It has been thought that adding these variables which have definite values all the time is enough to ensure the definiteness of measurement results and further solve the measurement problem.  
However, if the mental state cannot supervene on these additional variables (e.g. due to certain restrictions such as the no-signaling theorem), then even though these variables have definite values at all time, they are unable to account for our definite experience and thus do not help solve the measurement problem (see also Barrett, 2005). 

\section{Collapse theories}

I have argued that one will meet some difficulties if assuming the mental state of an observer supervenes either on certain branches of her wave function or on other additional variables. 
This seems to suggest that Everett's and Bohm's theories are not promising solutions to the measurement problem. 
Moreover, this also suggests that the mental state of an observer may supervene directly on her wave function, and 
collapse theories\index{collapse theories} may be in the right direction to solve the measurement problem\index{measurement problem}.\footnote{I will consider only objective versions of collapse theories here.} 

However, collapse theories\index{collapse theories} are still plagued by a few problems such as the tails problem\index{collapse theories!tails problem} (Albert\index{Albert, David Z.} and Loewer\index{Loewer, Barry}, 1996). In particular, it seems that the structured tails problem\index{collapse theories!tails problem!structured} has not been solved in a satisfactory way (see McQueen\index{McQueen, Kelvin J.}, 2015 and references therein).
The problem is essentially that collapse theories\index{collapse theories} such as the GRW\index{GRW theory} theory predicts that the post-measurement state is still a superposition of different outcome branches with similar structure (although the modulus squared of the coefficient of one branch is close to one), and they need to explain why high modulus-squared values are macro-existence determiners. 
In my view, the key to solving the structured tails problem\index{collapse theories!tails problem!structured} is not to analyze the connection between high modulus-squared values and macro-existence, but to analyze the connection between these values and our experience of macro-existence, which requires us to further analyze how the mental state of an observer supervenes on her wave function.\footnote{Note that this issue is independent of whether the observer can correctly report her mental content, which is related to the bare theory  (Albert, 1992; Barrett, 1999).}

Admittedly this is an unsolved, difficult issue. I will give a brief analysis here (see Gao, 2016 for a more detailed analysis). Consider an observer M being in the following superposition:

\begin{equation}
\alpha \ket{1}_P \ket{1}_M+\beta \ket{2}_P \ket{2}_M,
\label{sss}
\end{equation}

\noindent where $\ket{1}_P$ and $ \ket{2}_P$ are the states of a pointer being centered in positions $x_1$ and $x_2$, respectively, $\ket{1}_M$ and $\ket{2}_M$ are the physical states of the observer M who  observes the pointer being in positions $x_1$ and $x_2$, respectively, and $\alpha$ and $\beta$, which are not zero, satisfy the normalization condition $|\alpha|^2+|\beta|^2=1$.
The question is: What does M observe when she is physically in the above superposition?

First of all, it can be seen that the mental content of the observer M is related to the modulus squared of the amplitude of each branch of the superposition she is physically in.
When $|\alpha|^2$=1 and $|\beta|^2$=0,  M will observe the pointer being only in position $x_1$.
When $|\alpha|^2$=0 and $|\beta|^2$=1,  M will observe the pointer being only in position $x_2$.
When $\alpha = \beta =1/\sqrt{2}$, by the symmetry of the two branches the mental content of M will be neither the content of observing the pointer being in position $x_1$ nor the content of observing the pointer being in position $x_2$.


Next, it can be argued that the mental content of the observer M is also related to the phase of each branch of the superposition she is physically in.
Assume this is not the case. Then when $\alpha = -\beta =1/\sqrt{2}$ and when $\alpha = \beta =1/\sqrt{2}$, the mental content of M will be the same, which is neither the content of observing the pointer being in position $x_1$ nor the content of observing the pointer being in position $x_2$.
Then, when M is in a superposition of these two physical states, her mental content is still the same. 
However, since the superposition of these two states is $\ket{1}_P \ket{1}_M$, the observer M being in this superposition will observe the pointer being  in position $x_1$. 
This leads to a contradiction. 
Note that the mental content of M is related only to the relative phase of the two branches of the superposition she is physically in, since an overall phase has no physical meaning, and two physical states with only a difference of overall phase are in fact the same physical state. 

Now I will analyze how the mental content of the observer M is determined by the amplitude and phase of each branch of the superposition she is physically in. 
This is a difficult task. And I can only give a few speculations here.
Let us first see a few special cases. 
When $|\alpha|^2$=0 or $|\beta|^2=0$, the mental content of the observer M does not contain the content of observing the pointer being in position $x_1$ or $x_2$.
Similarly, the mental content of the observer M does not contain the content of observing the pointer being in another position $x_3$ which is different from  $x_1$ and  $x_2$, since the amplitude of the corresponding term $\ket{3}_P \ket{3}_M$ is exactly zero.
On the other hand, when $|\alpha|^2=1$ or $|\beta|^2=1$, the mental content of the observer M is the content of observing the pointer being in position $x_1$ or $x_2$. 
Then when $|\alpha|^2 \neq 0$ and $|\beta|^2 \neq 0$, the mental content of the observer M can only contain the content of observing the pointer being in position $x_1$ and the content of observing the pointer being in position $x_2$.
Moreover, according to the above analysis, how these two contents constitute the whole mental content of the observer M is determined by the values of $\alpha$ and $\beta$.

It seems relatively easy to conjecture how the modulus squared of the amplitude determines the mental content of the observer M.
Again, let us see a few special cases.
When $|\alpha|^2=0$, the mental content of the observer M does not contain the content of observing the pointer being in position $x_1$.
When $|\alpha|^2=1$, the mental content of the observer M contains only the content of observing the pointer being in position $x_1$.
Similarly, when $|\beta|^2=0$,  the mental content of the observer M does not contain the content of observing the pointer being in position $x_2$.
When $|\beta|^2=1$, the mental content of the observer M contains only the content of observing the pointer being in position $x_2$.
Then it seems reasonable to assume that  the mental property determined by the modulus squared of the amplitude is a certain property of vividness of conscious experience.
For example, when $|\alpha|^2$ is close to one the conscious experience of M observing the pointer being in position $x_1$ is the most vivid, while when $|\alpha|^2$ is close to zero, the conscious experience of M observing the pointer being in position $x_1$ is the least vivid.
In particular, when  $|\alpha|^2=|\beta|^2=1/2$, the conscious experience of M observing the pointer being in position $x_1$ and the conscious experience of M observing the pointer being in position $x_2$ have the same intermediate vividness.
However, it seems more difficult to conjecture the nature of the mental property determined by the relative phase. It is probably a new property which we don't know and have not experienced either. 

To sum up, I have argued that the mental content of an observer is related to both the amplitude and relative phase of each branch of the superposition she is physically in, and it may be composed of the mental content corresponding to every branch of the superposition. Moreover, the modulus squared of the amplitude of each branch may determine the vividness of the mental content corresponding to the branch. 

It can be seen that the above analysis of how the mental state of an observer supervenes on her wave function may help solve the structured tails problem of collapse theories. 
In particular, if assuming the modulus squared of the amplitude of each branch indeed determines the vividness of the mental content corresponding to the branch, then the structured tails problem may be solved. 
Under this assumption, when the modulus squared of the amplitude of a branch is close to zero, the mental content corresponding to the branch will be the least vivid.
It is conceivable that below a certain threshold of vividness an ordinary observer or even an ideal observer will not be consciously aware of the corresponding mental content.
Then even though in collapse theories the post-measurement state of an observer is still a superposition of different outcome branches with similar structure, the observer can only be consciously aware of the  mental content corresponding to the branch with very high amplitude, and the branches with very low amplitudes will have no corresponding mental content appearing in the whole mental content of the observer. 
This will solve the structured tails problem of collapse theories. 

\section{Conclusions}

It has been realized that the measurement problem is essentially the determinate-experience problem.  
The problem is not only to explain how the linear dynamics can be compatible with the appearance of definite measurement results obtained by physical devices, but also, and more importantly, to explain how the linear dynamics can be compatible with the existence of definite experiences of conscious observers. 
This suggests that in order to formulate and solve the measurement problem we need to analyze how the mental state of an observer relates to her physical state. 
However, the mental aspect of the measurement problem has been ignored in the conventional formulation of the problem, and the analysis of the forms of psychophysical connection in the solutions to the measurement problem seems not systematic and complete either in the literature. 

In this paper, I propose a new, mentalistic formulation of the measurement problem which lays more stress on the psychophysical connection.  
It is pointed out that the three main solutions to the measurement problem, namely Everett's theory, Bohm's theory and collapse theories, correspond to three different forms of psychophysical connection.
Moreover, I argue that the forms of psychophysical connection required by Everett's and Bohm's theories have potential problems, while an analysis of how the mental state of an observer supervenes on her wave function may help solve the structured tails problem of collapse theories. 
This seems to suggest that collapse theories may be in the right direction to solve the measurement problem.

\section*{Acknowledgments}
I am very grateful to two anonymous referees of this journal for their insightful comments, constructive criticisms and helpful suggestions. 
The basic idea of this paper came to my mind when I taught \emph{Philosophy of Quantum Mechanics} to the postgraduates at Chinese Academy University. I thank the International Conference Center of the University for providing comfortable accommodation. 
I am also grateful to Arthur Fine, Kelvin McQueen, Peter Lewis, Mark Stuckey, and Ken Wharton for helpful discussions at the 2016 International Workshop on Quantum Observers hosted by \emph{International Journal of Quantum Foundations}. 
This work is partly supported by a research project grant from Chinese Academy of Sciences and the National Social Science Foundation of China (Grant No. 16BZX021). 

\section*{References}
\renewcommand{\theenumi}{\arabic{enumi}}
\renewcommand{\labelenumi}{[\theenumi]}
\begin{enumerate}


\item Albert, D. Z. (1992). Quantum Mechanics and Experience. Cambridge, MA: Harvard University Press.

\item Albert, D. Z. and B. Loewer. (1988). Interpreting the Many Worlds Interpretation, Synthese, 77, 195-213.
\item Albert, D. Z. and B. Loewer (1996). Tails of Schr\"{o}dinger's Cat. In Perspectives on Quantum Reality, eds. R. Clifton. Dordrecht: Kluwer Academic Publishers.

\item Barrett, J. A. (1999). The Quantum Mechanics of Minds and Worlds. Oxford: Oxford University Press.
\item  Barrett, J. A. (2005). The preferred basis problem and the quantum mechanics of everything. British Journal for the Philosophy of Science 56 (2), 199-220.

\item Bedard, K. (1999). Material Objects in Bohm's Interpretation. Philosophy of Science 66, 221-242.

\item Bohm, D. (1952). A suggested interpretation of quantum theory in terms of ``hidden" variables, I and II. Physical Review 85, 166-193.

\item  Brown, H. R. (1996). Mindful of quantum possibilities. British Journal for the Philosophy of Science, 47, 189-200.
\item Brown, H. R. and D. Wallace (2005). Solving the measurement problem: de Broglie-Bohm loses out to Everett, Foundations of Physics 35, 517-540.

\item Butterfield, J. (1998). Quantum curiosities of psychophysics. In J. Cornwell (ed.), Consciousness and Human Identity, Oxford University Press, 122-157.

\item DeWitt, B. S. and N. Graham (eds.). (1973). The Many-Worlds Interpretation of Quantum Mechanics. Princeton: Princeton University Press.

\item Everett, H. (1957). `Relative state' formulation of quantum mechanics. Rev. Mod. Phys. 29, 454-462.

\item Gao, S. (2016). What does it feel like to be in a quantum superposition? http://philsci-archive.pitt.edu/11811/.

\item Gao, S. (2017). Failure of psychophysical supervenience in Everett's theory. In preparation.

\item  Ghirardi, G. C. (2011). Collapse Theories, The Stanford Encyclopedia of Philosophy (Fall 2008 Edition), Edward N. Zalta (eds.), http://plato.stanford.edu/archives/win2011/entries/qm-collapse/.

\item Lewis, P. J.  (2007a). How Bohm's Theory Solves the Measurement Problem. Philosophy of Science 74, 749-760.

\item Lewis, P. J.  (2007b). Empty Waves in Bohmian Quantum Mechanics. British Journal for the Philosophy of Science 58, 787-803.

\item Maudlin, T. (1995a). Three measurement problems. Topoi 14, 7-15.

\item Maudlin, T. (1995b). Why Bohm's theory solves the measurement problem. Philosophy of Science 62, 479-483.

\item Maudlin, T. (2007). Completeness, supervenience, and ontology. Journal of Physics A: Mathematical and Theoretical, 40, 3151-3171.

\item McLaughlin, B. and Bennett, K. (2014). "Supervenience", The Stanford Encyclopedia of Philosophy (Spring 2014 Edition), Edward N. Zalta (ed.), https://plato.stanford.edu/archives/spr2014/entries/ \\supervenience/.

\item  McQueen, K. J. (2015). Four tails problems for dynamical collapse theories. Studies in History and Philosophy of Modern Physics 49, 10-18.

\item  Stone, A. D. (1994). Does the Bohm theory solve the measurement problem?, Philosophy of Science  62, 250-266.

\item Valentini, A. (1992). On the Pilot-Wave Theory of Classical, Quantum and Subquantum Physics. Ph.D. Dissertation. Trieste, Italy: International School for Advanced Studies.

\item  Wallace, D. (2012). The Emergent Multiverse: Quantum Theory according to the Everett Interpretation. Oxford: Oxford University Press.

\item Zeh, H. D. (1981). The Problem of Conscious Observation in Quantum Mechanical Description, Epistemological Letters of the Ferdinand-Gonseth Association in Biel (Switzerland), 63. Also Published in Foundations of Physics Letters. 13 (2000) 221-233.

\item Zeh, H. D. (1999). Why Bohm's quantum theory?, Foundations of Physics Letters 12, 197-200.

\end{enumerate}
\end{document}